\documentclass[preprint,showpacs,preprintnumbers,amsmath,amssymb]{revtex4}

\usepackage{graphicx}
\usepackage{dcolumn}
\usepackage{bm}

\begin{document}

\preprint{APS/123-QED}

\title{Low-Light-Level Optical Interactions with Rubidium Vapor in a Photonic Bandgap Fiber}

\author{Saikat Ghosh}
\author{Amar R. Bhagwat}%
\author{C. Kyle Renshaw}
\author{Shireen Goh}
\author{Alexander L. Gaeta}%
\email{a.gaeta@cornell.edu}
\affiliation{%
School of Applied and Engineering Physics\\
Cornell University\\
Ithaca, NY 14853}%
\author{Brian J. Kirby}%
\affiliation{%
Sibley School of Mechanical and Aerospace Engineering \\
Cornell University\\
Ithaca, NY 14853}%

\newpage{}
\begin{abstract}
\par We show that a Rubidium vapor can be produced within the core of a photonic band-gap fiber yielding
an optical depth in excess of 2000. Our technique for producing the
vapor is based on coating the inner walls of the fiber core with an
organosilane and using light-induced atomic desorption to release Rb
atoms into the core.  We develop a model to describe the dynamics of
the atomic density, and as an initial demonstration of the potential
of this system for supporting ultra-low-level nonlinear optical
interactions, we perform electromagnetically-induced transparency
with control-field powers in the nanowatt regime, which represents
more than a 1000-fold reduction from the power required for bulk,
focused geometries.

\end{abstract}

\maketitle


\noindent Remarkable advances have been made in the past decade in
generating and controlling quantum states of light using atomic
vapors, including the realization of on-demand single-photon
sources~\cite{1}, manipulation of photonic states~\cite{2}, and
storage and retrieval of the states with high fidelity~\cite{3,4}.
Much of the motivation for these efforts has been to realize a
practical quantum network~\cite{5}.  In most cases, the underlying
optical process for realizing these schemes has been the phenomenon
of electromagnetically-induced transparency (EIT)~\cite{6} in which
a coherent superposition of atomic states is created by a strong
control field such that an optically thick atomic ensemble is
rendered transparent to a weak, resonant probe field. The concept of
EIT has been applied and expanded to schemes that allow two
extremely weak fields, which in principle can consist of single
photon pulses, to strongly interact~\cite{7,8,9,10}. Practical
implementation of these proposals in which a single photon can
switch another photon will lead to the realization of critical
components (e.g., a quantum phase gate) for quantum information
applications~\cite{11}.

\par The two generic requirements to achieve EIT-based, ultralow-level optical interactions are: 1) a large optical depth
$\kappa = nL\sigma$, where $n$ is the density of the atomic sample
of length $L$, and $\sigma$ is the atomic absorption cross-section,
and 2) confinement of the light beams to an area $A$ comparable to
the atomic scattering cross-section of
$3\lambda^2/2\pi$~\cite{8,9,10,12}. For example, the phase shift due
to nonlinear interactions between few photon-pulses in the proposed
scheme of Andr\'{e} et al.~\cite{10} and the inverse of the critical
power required to switch a signal field in the four-level scheme of
Harris and Yamamoto~\cite{8} are each proportional to $\kappa /A$.
In order to maximize the optical depth, a natural choice for the
atomic ensemble is an alkali atom due to its relatively simple
energy-level structure and its large $\sigma$ as compared to, for
example, molecules with ro-vibrational transitions~\cite{13}.  There
is a limit on how much $\kappa$ can be increased by increasing the
density $n$ since this will result in undesirable dephasing effects
due to atomic collisions.  Alternatively, increasing the length $L$
of the atomic sample can further enhance the optical depth. However,
in a bulk focused geometry, this length $L$ is limited to the
Rayleigh length, which can only be increased by a corresponding
increase in beam area  $A$~\cite{5,10}, and thus no change in the
quantity $\kappa/A$.

\begin{figure}
\includegraphics[width=3.2in]{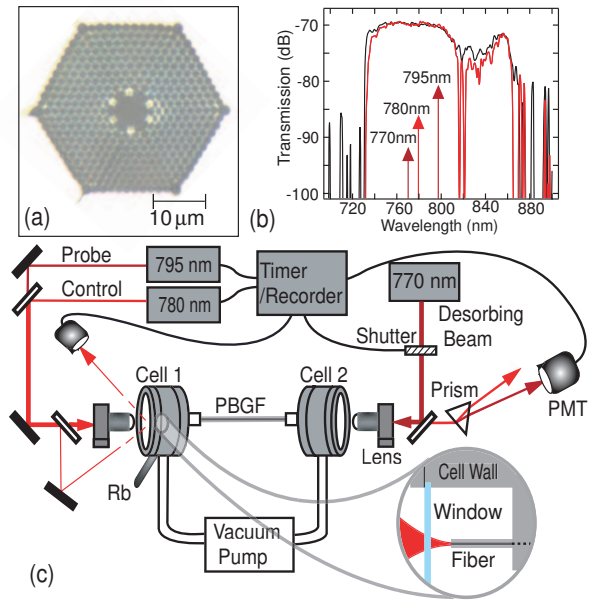}
\caption{(a) Transverse cross-section of the photonic band-gap fiber
(PBGF) used in the experiment. (b) Measured transmission band gap of
the PBGF before (black) and after (red) deposition of
octadecyldimethylmethoxysilane (ODMS) on the fiber walls. The arrows
indicate from left to right the wavelengths of the desorbing beam,
the control and the probe for EIT. (c) A schematic of the
experimental setup. Part of Cell 1 is expanded to illustrate the
region of the cell in front of the fiber tip.  The beam reflected
off a mirror at the back end of Cell 1 is used to calibrate the
density in the cell.}
\end{figure}
\par In this Letter, we realize a new experimental geometry for alkali vapors that overcomes the limitations of bulk
focused geometries by using light-induced atomic desorption (LIAD)
to produce controllable densities of Rb atoms within a suitably
coated core of a photonic band-gap fiber (PBGF).  The photonic
crystal structure~\cite{14} surrounding the core of a PBGF [see Fig.
1(a)] provides an unmatched ability to tightly confine light with a
gas in a region a few microns in diameter over meter-long distances,
and thus provides an ideal system to perform nonlinear optics at
extremely low light levels~\cite{10,12}.  Recent demonstrations of
nonlinear optical processes in PBGFs using gases with relatively
weak nonlinearities has led to the observation of low-threshold
stimulated Raman scattering~\cite{15}, high-power soliton
generation~\cite{16}, and coherent resonant interactions with
molecules~\cite{13,16a}. By using a gas such as an alkali vapor with
a strong nonlinear optical response, such a fiber system can form a
the basis for creation, manipulation, storage and transmission of
photonic states in a fiber geometry.  However up until now, the
ability to inject alkali atoms into PBGFs has eluded researchers due
to the strong interaction of the vapor with silica walls of the
fiber.
\par The primary challenge to creating a useful vapor of Rb atoms within a PBGF is that Rb vapor attacks and adheres
to silica glass fiber walls through both physisorption, in which the
atoms stick to the surface for a finite time, and through
chemisorption, in which the atoms are lost to the wall~\cite{17}.
These issues are particularly severe for a fiber with a core
6-$\mu$m in diameter~\cite{18}, which is a factor of $10^8$ smaller
than the atomic mean free path at room temperature. Furthermore, for
the fraction of atoms undergoing physisorption, the spin-decoherence
is large~\cite{19}, which makes this system unsuitable as a
practical quantum device. However, treating a glass surface with
paraffin or siloxane coatings~\cite{20} significantly alters the
Rb-surface interaction properties such that the wall-induced
dephasing rate decreases by four orders of magnitude with a
significant reduction in chemisorption when compared with that for
uncoated silica glass walls~\cite{21}. In addition, atoms attached
to such coated walls can be released by sudden exposure to optical
radiation through the non-thermal process termed LIAD~\cite{22, 23,
24}.

 \par We applied these techniques to the fiber geometry by surface-modifying the core walls of a PBGF
 \{AIR-6-800, Crystal Fibre, with a core diameter of 6 $\mu$m [Fig.1(a)] and a bandgap extending from 750 to 810 nm,
 chosen to accommodate the D1 and D2 lines of Rb at 795 nm and 780 nm, respectively [Fig.1(b)]\} with a monolayer
 of $C_{18}H_{35}$ moieties by self-assembly of octadecyldimethylmethoxysilane (ODMS)~\cite{25} via
 hydrolysis and condensation from solution. This monolayer deposition technique avoids the clogging of the fiber core
 that would occur for vacuum deposition of paraffin. The coating solution was introduced from polyolefin syringes to
 the PBGF via swaged PEEK fixtures, , incubated in the core to facilitate the monolayer deposition, and then flushed out. As seen in Fig.1(b), the bandgap of the fiber is preserved
 following the coating process. Following the monolayer deposition, the ends of the PBGF of length $L_{fib} = 25$ cm are
 placed in separate vacuum cells which are each connected to an ultra-high vacuum system [Fig. 1(c)]. One of the cells
 [left cell, in Fig. 1(c)] is subsequently exposed to natural Rb vapor, at a pressure of $10^{-6}$ Torr, and a beam from an
 external-cavity diode laser reflected from a mirror inside the cell allows for monitoring the Rb density $n_0$ in the cell,
 which is kept at $n_0 =2.1\times 10^{10}$ cm$^{-3}$. Bulk condensation of Rb vapor inside the fiber core is prevented by
 maintaining a constant temperature along the entire length of the fiber.  In the absence of LIAD, a finite fraction of
 atoms diffuse a length $z$ down the fiber core of radius $r_{fib}$, and this atomic flux, which is proportional to the thermal
 velocity, can be estimated in the Knudsen limit~\cite{26}.  A steady-state condition, known as "ripening", is reached
 when this flux equals the rate of adsorption to the wall surface at $z$~\cite{24}. For the PBGF in which
 $L_{fib}/r_{fib}\simeq
 10^5$, this ripening time can be extremely large. For simplicity, we rely on the Knudsen
 flow for the atoms to diffuse down the core. However, the atomic flux can
be significantly enhanced with techniques such as
 light-induced drift~\cite{27} or dipole-force guidance~\cite{18} into the core. The total number of atoms in the core is determined by monitoring the
 transmission of a weak laser beam coupled to the core and by scanning over the $D_1$ transition. We take into account the density
 contribution due to the beam path in the cell before the fiber [Fig. 1(c)] and fit the resulting absorption trace to the
 transmission coefficient,
\begin{equation}
T(\omega)=exp\left\{-{\int_{0}}^{L_{fib}}n(z)dz{\int_{-\infty}^{\infty}}\sigma(v,\gamma,\omega)W(v)dv\right\},
\end{equation}
where $n(z)$ is the atomic density at position $z$ in the core, and
the atomic cross-section $\sigma(v,\gamma,\omega)$ of the transition
is a function of the atomic velocity $v$, the homogeneous linewidth
$\gamma$, and the laser frequency $\omega$. The cross-section is
averaged over the Doppler profile $W(v)$. From the fit of the
experimental trace to Eq. 1, we estimate that in the absence of LIAD
the total number of atoms in the core to be
$N_{core}=A_{fiber}{\int_{0}}^{L_{fib}}n(z)dz\simeq 1.93\times 10^3$
, where $A_{fiber}$ is the cross-sectional area of the fiber core.

\begin{figure}
\includegraphics[width=3.0in]{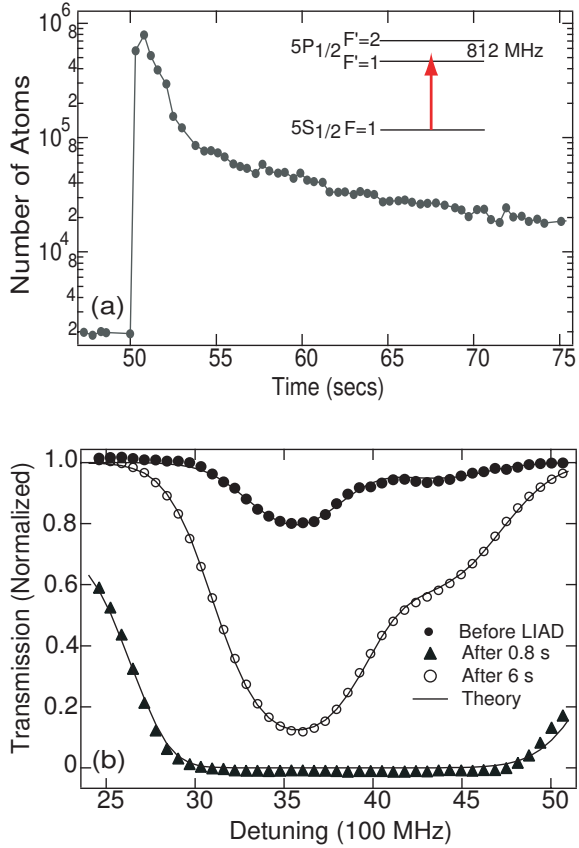}
\caption{(a) Measurement of the number of Rb atoms in the core of
the photonic band-gap fiber as a function of time, obtained from a
theoretical fit to the transition shown in the inset. (b) Variation
of the absorption lineshapes at three different times. The
corresponding theoretically predicted absorption profile is shown by
the solid line.}
\end{figure}

\par In our experiments, the intensity of the probe beam is maintained at 100 pW, which is an order of magnitude
lower than the measured saturation power of 3 nW in the core. By
coupling a desorbing beam counterpropagating to the probe beam  into
the core and tuned far off resonance at 770 nm, we observe a
dramatic increase in the total number of atoms. Figure 2(a) shows
that the atomic population undergoes a nearly instantaneous increase
by three orders of magnitude after the turn-on of the 1-mW desorbing
beam, with a maximum optical depth in excess of 2000. While a recent
experiment~\cite{28} has reported Rb desorption from porous silica,
we have not observed measurable desorption in uncoated fibers.  From
the fitted absorption profile, we estimate a homogeneous linewidth
of $\gamma$ = 96 MHz for the $F=1\rightarrow F'=1,2$  transitions.
The broadening associated with the dipole dephasing of the atoms
colliding with core wall can be estimated from the wall-collisional
frequency $\bar{v}/2r_{fib} \simeq  $85 MHz, where $\bar{v}$ is the
thermal velocity, which suggests that the dominant contribution to
the homogeneous linewidth is due to the wall-collisional dephasing
as opposed to atomic collisional broadening. This is consistent with
our data that shows the linewidth to be nearly constant as a
function of time as the density varies over two orders of magnitude.

\begin{figure}
\includegraphics[width=3.0in]{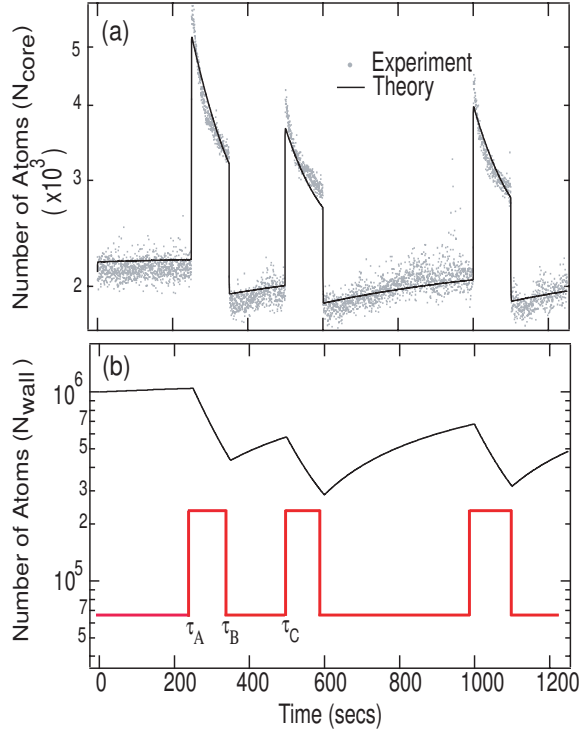}
\caption{(a) Measurement of the number of Rb atoms in the core of
the photonic band-gap fiber (dots) as a function of time in the
presence of a 60-$\mu$W desorbing beam at 770 nm, with the
theoretical fit (line) from Eq.(2) to the experiment. (b)
Corresponding variation in the number of atoms adsorbed to the wall;
the line in red illustrates the timing for the desorbing beam.}
\end{figure}

\par As a result of the transient nature of the density of atoms within the core, application of this fiber system to
low-light-level nonlinear optics requires a basic understanding of
the dynamics of the desorbed atoms in order to determine the time
window in which useful interactions can be performed. We apply the
following simple model~\cite{23, 24} to explain most of the features
of the observed density dynamics. The temporal evolution of the
total number $N_{core}$ of atoms in the core is modeled by the
equation,
\begin{align}
\frac{dN_{core}}{dt} = -\frac{\rho\bar{v}A_{fiber}}{4}N_{core} +
(\gamma_{T} + aI_d)N_{wall} \\ \nonumber + \xi(N_{cell} - N_{core}).
\end{align}
The first term on the right-hand-side is the rate of loss of atoms
from the core due to collisions with the walls, with the probability
$\rho$  that an atom sticks to the wall. The second term is the
contribution from the $N_{wall}$ atoms stuck to the core walls. This
increase consists of two contributions: a light-independent thermal
desorption rate $\gamma_{T}$ and a light-induced desorption rate,
which is proportional to the intensity $I_d$ of the desorbing beam
at the core wall with a proportionality constant $a$. It is via this
second process that the atomic density and the optical depth of the
system can be varied. The last factor in Eq. 2 represents the
relaxation of the atomic number to the steady-state value  $N_{cell}
= n_0A_{fib}L_{fib}$ at a rate $\xi$. An equation similar to Eq. 2
is assumed for $N_{wall}$, where the first two factors contribute
with opposite signs, leading to an increase and decrease of atoms
stuck to the walls, respectively. Fig. 3(a) shows a plot of
variation in the atomic population for various exposures of the
desorbing beam of $60 \mu W$, together with a fit of the adopted
model. The model was fitted to the experiment up to time $\tau_B$
[Fig. 3(b)], and the subsequent comparison of the theory to the
experiment shows excellent agreement.
\par To demonstrate the potential of this system to facilitate nonlinear optics at low-light levels, we investigate
EIT in a $V$-type system with a probe at 100 pW and with control
powers as low as 10 nW; these control powers are smaller by more
than a factor of 1000X than what is typically used to achieve EIT in
bulk geometries~\cite{29}.  The probe is tuned to the  $F=2
\rightarrow F'=1$ transition of the $D_1$ line of $^{87}Rb$ at 795
nm, and a control field, copropagating with the probe, is tuned to
the $F=2 \rightarrow F'=3$  transition of the $D_2$ line at 780 nm.
For this particular level scheme [inset of Fig. 4(b)], optical
pumping between the hyperfine levels is avoided, since the $F=1
\rightarrow F'=3$ transition is dipole forbidden.  Furthermore, the
probe-field saturation and the optical pumping between the magnetic
sublevels tend to cancel each other, and as a result the observed
transparency is primarily due to pure EIT~\cite{29}.  To analyze
this system, we solve the density-matrix equations for a 3-level $V$
system in steady state, with level $a$ as the ground state and with
$b$ and $c$ as the two excited states [Fig. 4(b)]. The coherence
$\sigma_{ca}$ to first order in the probe field is given
by~\cite{13},

\begin{align}
\sigma_{ca}=
&\frac{-i\Omega_p}{2[\gamma_{ac}-i\delta_p+\frac{|\Omega_c|^2/4}{\gamma_{bc}+i(\delta_c
- \delta_p)}]} \hspace{3mm}\times \\ \nonumber &\left\{(\rho_{cc}^0
- \rho_{aa}^0)-\frac{|\Omega_c|^2(\rho_{bb}^0 -
\rho_{aa}^0)}{4(\gamma_{bc} + i\delta_c)[\gamma_{bc} + i(\delta_c -
\delta_p)]}\right\},
\end{align}
where  $\Omega_c$ ($\Omega_p$) and $\delta_c$ ($\delta_p$) are the
Rabi frequency and detuning, respectively, for the control (probe)
field, $\rho_{ii}^0, (i = a,b,c)$ are the steady-state population
distributions, $\gamma_{ij} = (\gamma_i + \gamma_j)/2 +
\gamma_{ij}^{coll}, (i,j = a,b,c)$ are the dephasing rates, and
$\gamma_i$ is the decay rate of level $i$. The imaginary part of the
Doppler-averaged susceptibility, calculated from this coherence, is
integrated over the length of the fiber to fit to the transmission
trace of the probe field.
\par Figure 4 shows results in which a 1-mW desorbing beam releases Rb atoms into the core, and a series of probe
transmission spectra are taken at ensuing time intervals. The time
(250 ms) to obtain a trace is chosen to be long compared to the
atomic time scales (100's ns) but short compared to the time scales
(secs) associated with the desorption dynamics of the atomic density
in the core.  The input power of the probe field is set to 100 pW,
and that of the control field is varied from 10 nW to 3 $\mu$W.
Figure 4(a) shows a typical trace of the probe field transmission in
presence of a 361-nW control field, together with the corresponding
theoretical fit as calculated from Eq. (3). Using the fitting
procedure described in~\cite{13}, we estimate a decay rate for the
coherence between the two upper states to be $\gamma_{bc}$ = 24 MHz
and the two-level decoherence rates $\gamma_{ai},(i = b,c)$ to be
between 90-100 MHz. Figure 4(b) shows the measured transparency
full-width at half maximum (FWHM) together with the corresponding
FWHM calculated from Eq. 3. The error bar denotes the variation in
measurements which were taken at different time intervals of atomic
desorption [ Fig 2(a)]. At higher powers or for transitions in which
optical pumping and saturation effects contribute, larger than 90\%
transparencies are observed [see inset of Fig. 4(a)].
\begin{figure}
\includegraphics[width=3.0in]{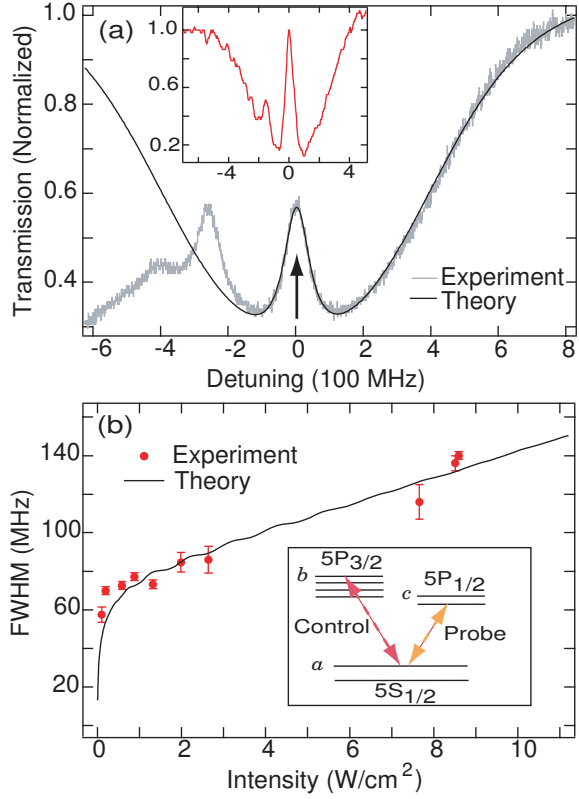}
\caption{(a) Transmission spectra of the probe field in presence of
a 361-nW control field. The arrow shows the transparency window due
to electromagnetically-induced transparency (EIT) together with the
corresponding theoretical plot fitted. The peak to the left of the
arrow corresponding to detuned $5S_{1/2},F=2\rightarrow
5P_{3/2},F'=2$ transition is not taken into account in the fit. The
inset shows transparency larger than 90\% for a probe scanned over
$5S_{1/2},F=1\rightarrow 5P_{1/2},F'=1$ with a 2.65-$\mu$W control
field tuned to $5S_{1/2},F=1\rightarrow 5P_{3/2},F'=1$  transition.
(b) Experimental (red) and theoretical (grey) variation of the EIT
linewidth as a function of the control intensity. The inset shows
the energy-level scheme for this system.}
\end{figure}
\par In conclusion, we have demonstrated a new technique to create a significant density of Rubidium vapor in the core
of a PBGF and as a proof of concept, we have demonstrated EIT in
this system with a control power as low as 10 nW which represents
more than a factor of $10^7$ reduction, as compared to
acetylene-based EIT in PBGF [13, 17]. Such a system represents a new
experimental geometry for performing nonlinear optics at extremely
low light levels due to its unmatched combination of strong light
confinement and long interaction lengths with atoms of large optical
cross-section.
\par We thank D. Gauthier, J. E. Sharping, K. D. Moll, K. Koch for stimulating discussions and D. Gauthier for the
loan of the photomultiplier tube. We gratefully acknowledge support by the Center for Nanoscale Systems, supported
by the NSF under Grant No. EEC-0117770, the Air Force Office of Scientific Research under Contract No. F49620-03-0223,
and DARPA under the Slow-Light program.

{}

\end{document}